\documentclass[iop]{emulateapj}
\newcommand{\sone}{$\mu +\sigma$}
\newcommand{\stwo}{$\mu +2\sigma$}
\usepackage{amsmath}

\begin{document}
\title{Of Genes and Machines: application of a combination of machine
  learning tools to astronomy datasets} \shorttitle{Of Genes and
  Machines} \shortauthors{S. Heinis et al.}

\author{S. Heinis\altaffilmark{1}, S. Kumar\altaffilmark{1}, S. Gezari\altaffilmark{1}, W. S. Burgett\altaffilmark{2}, K. C. Chambers\altaffilmark{2}, P. W. Draper\altaffilmark{3}, H. Flewelling\altaffilmark{2}, N. Kaiser\altaffilmark{2}, E. A. Magnier\altaffilmark{2}, N. Metcalfe\altaffilmark{3}, C. Waters\altaffilmark{2}}

\altaffiltext{1}{Department of Astronomy, University of Maryland, College Park, MD, USA}
\altaffiltext{2}{Institute for Astronomy, University of Hawaii at Manoa, Honolulu, HI 96822, USA}
\altaffiltext{3}{Department of Physics, Durham University, South Road, Durham DH1 3LE, UK}

\begin{abstract}
We apply a combination of a Genetic Algorithms (GA) and
Support Vector Machines (SVM) machine learning algorithm to solve two
important problems faced by the astronomical community: star/galaxy
separation, and photometric redshift estimation of galaxies in survey
catalogs. We use the GA to select the relevant features in the first
step, followed by optimization of SVM parameters in the second step to
obtain an optimal set of parameters to classify or regress, in process
of which we avoid over-fitting. We apply our method to star/galaxy
separation in Pan-STARRS1 data. We show that our method correctly
classifies 98\% of objects down to $i_{\rm P1}= 24.5$, with a
completeness (or true positive rate) of 99\% for galaxies, and 88\%
for stars. By combining colors with morphology, our
star/classification method yields better results than the new
SExtractor classifier \texttt{spread\_model} in particular at the
faint end ($i_{\rm P1}>22$). We also use our method to derive
photometric redshifts for galaxies in the COSMOS bright
multi-wavelength dataset down to an error in $(1+z)$ of
$\sigma=0.013$, which compares well with estimates from SED fitting on
the same data ($\sigma=0.007$) while making a significantly smaller
number of assumptions.

\end{abstract}

\section{Introduction}
Astronomy has witnessed an ever-increasing deluge of data over the
past decade. Future surveys will gather very large amounts of data
daily that will require on-the-fly analysis to limit the amount of
data that can be stored or analyzed, and allow timely discoveries of
candidates to be followed-up: for instance Euclid \citep[][100 GB per
  day]{Laureijs_2011}, WFIRST \citep[][1.3TB per day]{Spergel_2015},
or the Large Synoptic Survey Telescope \citep[LSST, ][10 TB per
  day]{Ivezic_2008}. The evolution of the type, volume, cadence of the
astronomical data requires the advent of robust methods that will
enable a maximal rate of extraction of information from the data. This means
that for a given problem, one needs to make sure all the relevant
information has is made available in the first step, followed by the
use of a suitable method that is able to narrow down on the important
aspects of the data. This can be both for feature selection as well as
for noise filtering.

In machine learning parlance, the tasks required can be designated as
classification tasks (derive a discrete value: star vs galaxy for
instance) or regression tasks (derive a continuous value: photometric
redshift for instance.) Methods for classification or regression
usually are of two kinds: physically motivated, or
empirical\footnote{Formalisms that blend physically motivated and
  empirical have been proposed, see
  e.g. \citet{Budavari_2009}.}. Physically motivated methods use
templates built from previously observed data, like star or galaxy
Spectral Energy Distributions (SEDs) for instance in the case of
determining photometric redshifts. They also attempt at including as
much knowledge as we have of the processes involved in the problem at
stake, such as prior information. Physically motivated methods seem
more appropriate than empirical, however the very fact that they
require a good knowledge of the physics involved might be a important
limitation. Indeed, our knowledge of a number of processes involved in
shaping the SEDs of galaxies for instance is still quite limited
\citep[e.g.][ and references therein]{Conroy_2010}: whether is it
about the processes driving star formation, dust attenuation laws,
initial mass function, star formation history, AGN contribution etc \ldots
Hence choices need to be made: either make a number of assumptions, in
order to reduce the number of free parameters. Or, one can decide to
be more inclusive and add more parameters, but at the cost of
potential degeneracies. Physically motivated methods are also usually
limited in the way they treat the correlation between the properties
of the objects, because the only correlations taken into account are
those included in the models used, and might not reflect all of the
information contained in the data.

On the other hand, empirical methods require few or no assumptions
about the physics of the problem. The goal of an empirical method is
to build a decision function from the data themselves. The quality of
the generalization of the results to a full dataset depends of course
on the representativity of the training sample. We note however that
the question of the generalization also applies to physically
motivated methods, as they are also validated on the training data.

Depending on the methods used, transformations may need to be effected
before the method is applied; for example in the case of Support
Vector Machines \citep[SVM, e.g.][]{Boser_1992, Cortes_1995,
  Vapnik_1995, Smola_1998, Vapnik_1998, Duda_2001} linear separability is presumed, thereby
requiring non-linearly separable or regressible data to be kernel
transformed to enable this.

A challenge with empirical methods is that with growing numbers of
input parameters, it becomes prohibitive in terms of CPU time to use
all of them. It can also be counter-productive to feed the machine
learning tool with all of these parameters, as some of them might
either be too noisy, or not bring any relevant information to the
specific problem to tackle. Moreover, high dimensional problems suffer
from the bane of being over fit by machine learning methods
\citep{Cawley_2010}, thereby yielding high-dimensional non-optimal
solutions.  This requires sub-selection of relevant variables from a
large $N-$dimensional space (with $N$ potentially close to 1000). This
task itself can also be achieved using machine learning tools. Here we
present, to our knowledge, the first application to astronomy of the
combination of two machine learning techniques: Genetic Algorithms
\citep[GA, e.g.][]{Steeb_2014} for selecting relevant
features followed by SVM to build a decision/reward function. GA alone
have already been used in astronomy for instance to study the orbits
of exoplanets, and the SEDs of young stellar objects
\citep{Canto_2009}, the analysis of SNIa data \citep{Nesseris_2011},
the study of star formation and metal enrichment histories of resolved
stellar system \citep{Small_2013}, the detection of globular clusters
from HST imaging \citep{Cavuoti_2014}, or photometric redshifts
estimation \citep{Hogan_2015}. On the other hand, SVM have been
extensively used to solve a number of problems such as object
classification \citep{Zhang_2004}, the identification of red variables
sources \citep{Wozniak_2004}, photometric redshift estimation
\citep{Wadadekar_2005}, morphological classification
\citep{Huertas_2011}, and parameters estimation for Gaia
spectrophotometry \citep{Liu_2012}. We note also that the combination
of GA and SVM has already been used in a number of fields such as
cancer classification \citep[e.g.][]{Huerta_2006, Alba_2007},
chemistry \citep[e.g.][]{Fatemi_2007}, bankruptcy prediction
\citep[e.g.][]{Min_2006} etc \ldots We do not attempt here to provide
the most optimized results from this combination of methods. Rather,
we present a proof-of-concept that shows that GA and SVM yield
remarkable results when combined together as opposed to using the SVM
as a standalone tool.

In this paper, we focus on two tasks frequently seen in large surveys:
star/galaxy separation and determination of photometric redshifts of
distant galaxies. Star/galaxy separation is a classification
  problem that has been usually constrained purely from the morphology
  of the objects \citep[e.g.][]{Kron_1980, Bertin_1996,
    Stoughton_2002}. On the other hand, a number of studies used the
  color information with template fitting to separate stars from
  galaxies \citep[e.g.][]{Fadely_2012}. On top of these two common
  approaches, good results have been obtained by feeding morphology
  and colors to machine learning techniques
  \citep[e.g.][]{Vasconcellos_2011, Saglia_2012, Kovacs_2015}. Our
goal here is to extend the star/galaxy separation to faint magnitudes
where morphology is not as reliable. We apply here the combination
of GA with SVM to star/galaxy separation in the Pan-STARRS1 (PS1)
Medium Deep survey \citep{Kaiser_2010, Tonry_2012, Rest_2014}.

On the other hand, determining photometric redshifts is a well-studied
problem of regression that has been dealt with using a variety of
methods, including template fitting \citep[e.g.][]{Arnouts_1999,
  Benitez_2000, Brammer_2008, Ilbert_2009}, cross correlation
functions \citep{Rahman_2015}, Random Forests \citep{Carliles_2010},
Neural Networks \citep{Collister_2007}, polynomial fitting
\citep{Connolly_1995}, symbolic regression
\citep{Kron_Martins_2014}. We apply the GA-SVM to the zCOSMOS bright
sample \citep{Lilly_2007}.

The outline of this paper is as follows. In Sect. \ref{sec_data} we
present the data we apply our methods to, and also the training sets
we use: PS1, and zCOSMOS bright. In Sect. \ref{sec_methods} we briefly
describe our methods; we include a more detailed description of SVM in
an Appendix. We present our results in Sect. \ref{sec_results}, and
finally list our conclusions.

\section{Data}\label{sec_data}
We perform star/galaxy separation in PS1 Medium Deep data (see
Sect. \ref{sec_ps1}), using a training set built from COSMOS ACS
imaging (see Sect. \ref{sec_cosmos_sg}). We then derive photometric
redshifts for the zCOSMOS bright sample (see
Sect. \ref{sec_cosmos_z}).

\subsection{Pan-STARRS1 data}\label{sec_ps1}
The Pan-STARRS1 survey \citep{Kaiser_2010} surveyed 3/4 of the
northern hemisphere ($\delta>-30$\,deg), in 5 filters: $g_{\rm P1}$,
$r_{\rm P1}$, $i_{\rm P1}$, $z_{\rm P1}$, and $y_{\rm P1}$
\citep{Tonry_2012}. In addition, PS1 has obtained deeper imaging in 10
fields (amounting to 80 deg$^2$) called as the Medium Deep Survey
\citep{Tonry_2012, Rest_2014}. We use here data in the MD04 field,
which overlaps the COSMOS survey \citep{Scoville_2007}.

We use our own reduction of the PS1 data in the Medium Deep Survey. We
use the image stacks generated by the Image Processing Pipeline
\citep{Magnier_2006}, and also the CFHT $u$ band data obtained by
E. Magnier as follow up of the Medium Deep fields. We have at hand 6
bands: $u_{\rm CFHT}$, $g_{\rm P1}$, $r_{\rm P1}$, $i_{\rm P1}$,
$z_{\rm P1}$, and $y_{\rm P1}$. We perform photometry using the
following steps; we consider the PS1 \textit{skycell} as the smallest
entity: \textit{i)} resample \citep[using
  \texttt{SWarp},][]{Bertin_2002} the u band images to the PS1
resolution, and register all images; \textit{ii)} for each band fit
the PSF to a Moffat function, and match that PSF to the worse PSF in
each skycell; \textit{iii)} using these PSF-matched images, we derive
a $\chi^2$ image \citep{Szalay_1999}; \textit{iv)} we perform
photometry using the \texttt{SExtractor} dual mode
\citep{Bertin_1996}, detecting objects in the $\chi^2$ image and
measuring the fluxes in the PSF matched images: the Kron-like
apertures are defined from the $\chi^2$ image and hence are the same
over all bands ; \textit{v)} for each detection we measure
\texttt{spread\_model} in each band on the original, non PSF-matched
images. We consider here Kron magnitudes, which are designed to
contain $\sim 90\%$ of the source light, regardless of being a point
source star or an extended galaxy.

\texttt{spread\_model} \citep{Soumagnac_2013} is a discriminant
between the local PSF, measured with \texttt{PSFEx}
\citep{Bertin_2011} and a circular exponential model of scale length
$FWHM/16$ convolved with that PSF. This discriminant has been shown to
perform better than SExtractor previous morphological classifier,
\texttt{class\_star}.

\subsection{COSMOS data}

\subsubsection{Star/galaxy classification training set}\label{sec_cosmos_sg}
We use the star/galaxy classification from \citet{Leauthaud_2007}
derived from high spatial resolution Hubble Space Telescope (HST)
imaging (in the $F814W$ band) as our training set for star/galaxy
classification. They separate stars from galaxy based on their
\texttt{SExtractor} \texttt{mag\_auto} and \texttt{mu\_max} (peak
surface brightness above surface level). By comparing the derived
stellar counts with the models of \citet{Robin_2007},
\citet{Leauthaud_2007} showed that the stellar counts are in excellent
agreement with the models for $20<F814W<25$. We perform here
star/galaxy separation down to $i_{\rm P1}=24.5$.

\subsubsection{Spectroscopic redshift training set}\label{sec_cosmos_z}
We use data taken as part of the COSMOS survey
\citep{Scoville_2007}. We focus here on objects with spectroscopic
redshifts obtained during the zCOSMOS bright campaign
\citep[][$i<22.5$]{Lilly_2007}. We use the photometry obtained in 25
bands by various groups in broad optical and near-infrared bands:
\citet[][$u^*$,$B_J$,$V_J$,$g^+$, $r^+$,
  $i^+$,$z^+$,$K_s$]{Capak_2007}; narrow bands:
\citet[][$NB816$]{Taniguchi_2007} and \citet[][$NB711$]{Ilbert_2009};
intermediate bands \citet[][$IA427$, $IA464$, $IA505$, $IA574$,
  $IA709$, $IA827$, $IA484$, $IA527$, $IA624$, $IA679$, $IA738$,
  $IA767$]{Ilbert_2009}. We also use IRAC1 and IRAC2 photometry
\citep{Sanders_2007}. We consider here only objects with good redshift
determination (confidence class: 3.5 and 4.5),$z<1.5$ (which yields
5,807 objects) and measured magnitudes for all the bands listed above,
which finally leaves us with 5093 objects. We do not use objects with
other confidence class values, as the spectroscopic redshifts can be
erroneous in at least 15\% of cases (Ilbert, Private communication).

\section{Methods}\label{sec_methods}
We present here the two machine learning methods we use in this work:
Genetic Algorithms are used to select the relevant features, and
Support Vector Machines are used to predict the property of interest
using these features.

\subsection{Genetic Algorithms}
Genetic algorithms \citep[GA, e.g.][]{Steeb_2014} apply the
basic premise of genetics to the evolution of solution sets of a
problem, until they reach optimality. The definition of optimality
itself is debateable, however, the general idea is to use a reward
function to direct the evolution. That is, we evolve solution sets of
parameters (or genes) known as organisms through several generations,
according to some pre-defined evolutionary reward function. The reward
function may be a goodness of fit, or a function thereof that may take
into account more than just the goodness of fit. For example, one may
choose to determine subsets of parameters which optimize $\chi^2$ or
likelihood, Akaike Information criteria, energy functions, entropy,
etc.  The lengths of the first generation of organisms is chosen to
range from the minimum expected dimensionality of the parameter space
(which can be 1) to the maximum expected dimensionality of the
covariance.

The fittest organisms in a particular generation are then given higher
probabilities of being chosen to be the parents for successive
generations using a roulette selection method based on their
fitnesses. Parents are then cross-bred using a customized procedure -
in our method we first choose the lengths of the children to be based
on a tapered distribution where the taper depends on the fitnesses of
the parents. That way the length of the child is closer to that of the
fitter parent. The child is then populated using the genes of both
parents, where genes that are present in both parents are given twice
the weight as genes that are only present in one parent. The idea is
that the child should, with a greater probability, contain genes that
are present in both parents as opposed to the ones contained in only
one of them. This process iteratively produces fitter children in
successive generations and is terminated when no further improvement
in the average organism quality is seen. Like in biological genetics,
we also introduce a mutation in the genes with constant
probability. The genes which are chosen to be mutated, are replaced
with any of the genes that are not part of either parent, with uniform
probability of choosing from the remaining genes. This allows for
genes that are not part of the current gene pool to be expressed. The
conceptual simplicity of genetic algorithms combined with their
evolutionary analogy as applied to some of the hardest multi-parameter
global optimization problems makes them highly sought after.

\subsection{Support Vector Machines}

Support vector machines (SVM) \citep[e.g.][]{Boser_1992, Cortes_1995,
  Vapnik_1995, Smola_1998, Vapnik_1998, Duda_2001} is a machine learning
algorithm that constructs a maximum margin hyperplane to separate
linearly separable patterns. SVM is especially efficient in high
dimensional parameter spaces,where separating two classes of objects
is a hard problem, and performs with best case complexity
$\mathcal{O}n_{parameters}n^2_{samples}$. Where the data is not
linearly separable, a kernel transformation can be applied to map the
original parameter space to a higher dimensional feature space where
the data becomes linearly separable. For both problems we attempt to
solve in this paper, we use a Gaussian Radial Basis Function kernel,
defined as

\begin{equation}\label{eq_rbf}
K(x,x') =\exp(-\gamma ||x-x'||^2)
\end{equation}

Another advantage of using SVM is that there are established
guarantees of their performance which has been well documented in
literature. Also, the final classification plane is not affected by
local minima in the classification or regression statistic, which
other methods based on least squares, or maximum likelihood may not
guarantee. For a detailed description of SVM please refer to
\citet{Smola_1998, Vapnik_1995, Vapnik_1998}. We use here the
\texttt{python} implementation within the \texttt{Scikit-learn}
\citep{Pedegrosa_2011} module.

\subsection{Optimization procedure}\label{sec_optimization}

We combine the GA and SVM to select the optimal set of parameters that
enable one to classify objects or derive photometric redshifts. In
either case, we first gather all input parameters, and build color
combinations from all available magnitudes. We also consider here
transformations of the parameters, namely their logarithm and
exponential on top of their catalog values that we name 'linear'
afterwards, in order to capture non-linear dependencies on the
parameters. Note that any transformation could be used, we limited
ourselves to $\log$ and $\exp$ for sake of simplicity for this first
application. This way, the rate of change of the dependent parameter
as a function of the independent parameter in question is more
accurately captured, and dependencies on multiple transformations of
the same variable can also be captured. To eliminate scaling issues in
transformed spaces, we transform all independent parameters to between
-1 and 1.  The optimization is then performed the following two steps
iteratively until the end criterion or convergence criterion is met:
selection of the relevant set of features in the first, and automatic
optimization of SVM parameters to yield optimal
classification/regression given the parameters. For SVM
classification, the true positive rate is used as the fitness
function. For SVM regression, a custom fitness function based on the
problem at hand is chosen. For example, for SVM regressions on the
photometric redshift we use

\begin{equation}
\frac{1}{\sum_i \left(\frac{z^i_{phot} - z^i_{spec}}{z^i_{phot}}\right)^2}
\end{equation}

\begin{figure}[t]

\includegraphics[width=\hsize]{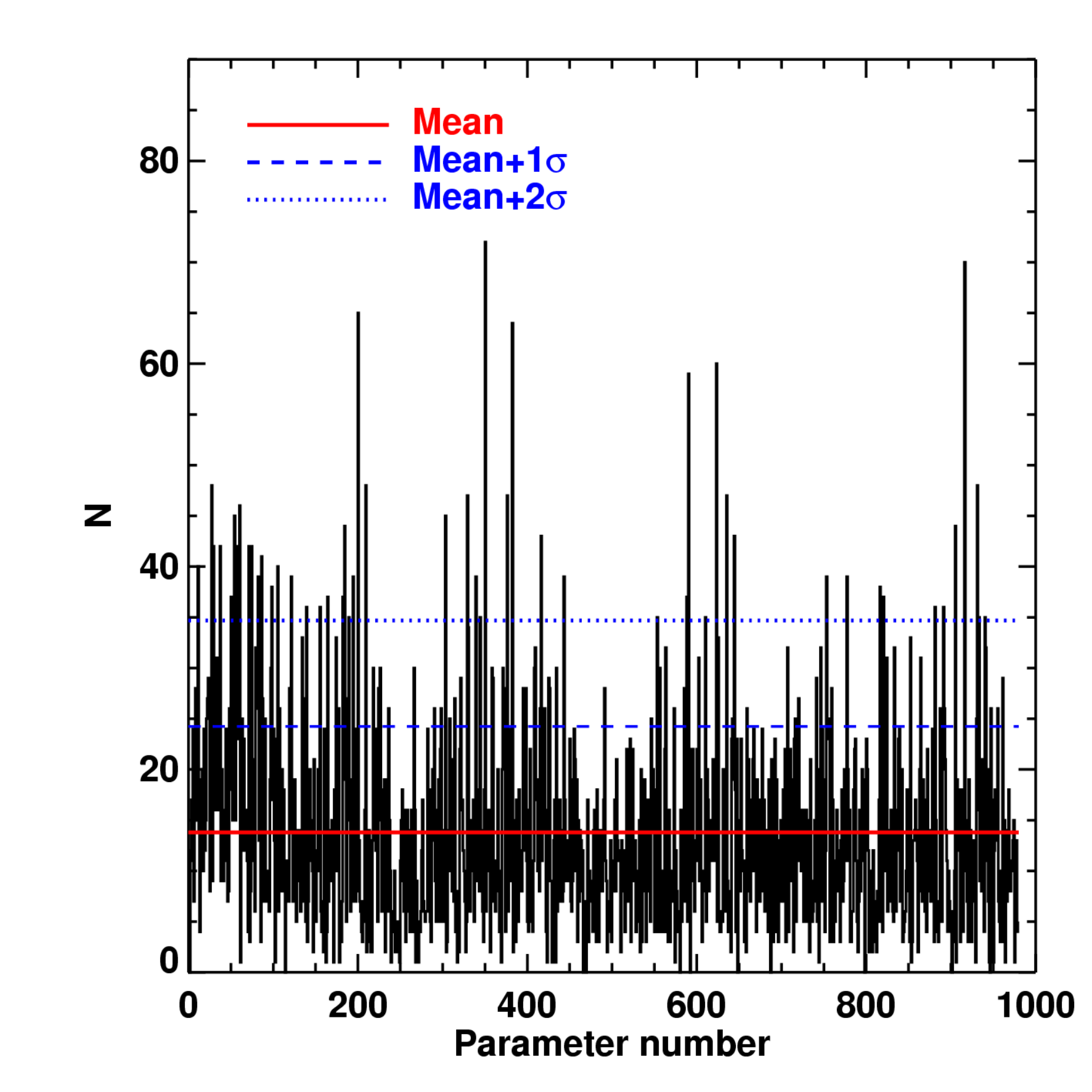} \figcaption{Example of
  posterior distribution for the parameters obtained from the
  GA-SVM. This posterior distribution was derived for the application
  to photometric redshifts (\S \ref{sec_photoz}). The red solid line
  shows the average occurence of the parameters; the dashed blue line
  the occurence at the mean plus one standard deviation, af the dotted
  blue line the occurence at the mean plus two standard deviations. We
  use here all parameters which appear more times than the mean plus
  one standard deviation in the posterior
  distribution. \label{fig_GA_posterior}}
\end{figure}

Once the fitness of all organisms has been evaluated, a new generation
of same size as that of the parent generation is created using
roulette selection. The GA then runs until it reaches a pre-defined
stopping criterion. We use here the posterior distribution of the
parameters. We stop the GA when all parameters have been used at least
10 times. We also use the posterior distribution to choose the optimal
set of parameters. Various schemes can be defined. For our
application, we restrict ourselves to characterizing the posterior
distribution by its mean $\mu$ and standard deviation $\sigma$ (see
Figure \ref{fig_GA_posterior}). We consider here all parameters that
appear more that $\mu+\sigma$ or $\mu+2\sigma$ times in the posterior
distribution, depending on which our results change significantly. For
instance, in the case of photometric redshifts the mean of the
posterior distribution is $\mu=13.8$, and the standard deviation
$\sigma=10.4$, we keep all parameters that occur more than 24 times in
the posterior distribution when using the $\mu+\sigma$ criterion.

\subsubsection{SVM parameters optimization}
SVM are not "black boxes'', but come with a well defined formalism,
and free parameters to be adjusted. For this application, we use the
$\nu$SVM version of the algorithm which allows to control the
fractional error and the lower limit on the fraction of support
vectors used. We use here $\nu=0.1$. In the case of classification
(star/galaxy separation), we are then left with only one free
parameter, the inverse of the width of the Gaussian Kernel, $\gamma$
(eq. \ref{eq_rbf}). In the case of regression (photometric redshifts),
we have an additional free parameter, the trade off parameter $C$ (see
Appendix \ref{app_svm}).

If not used with caution, machine learning methods can lead to
over-fitting : the decision function is biased by the training sample
and will not perform well on other samples.  In order to avoid
over-fitting and optimize the value of $\gamma$ and $C$, we perform
10-fold cross-validation: we divide the sample in 10 subsets; we then
perform classification or regression for each subset after training on
the 9 other subsets. We perform this cross validation for a grid of
$\gamma$ and $C$ values. In the case of SVM, the overfitting can be
measured by the fraction of objects used as support vectors, $f_{\rm
  SV}$. If $f_{\rm SV} \sim 1$, most of the training sample is used as
support vectors, will lead to poor generalization. For each
iteration, we also get $f_{\rm SV}$, in order to include it on our
cost function. For both applications, we minimize a custom cost
function that optimizes the quality of the classification or
regression, and $f_{\rm SV}$.

\section{Results}\label{sec_results}
\subsection{Star/Galaxy separation}
We use as inputs to the GA feature selection step all magnitudes
available for the PS1 dataset: $u_{\rm CFHT}$, $g_{\rm P1}$, $r_{\rm
  P1}$, $i_{\rm P1}$, $z_{\rm P1}$, and $y_{\rm P1}$; the
\texttt{spread\_model} values derived from each of these bands; the
ellipticity measured on the $\chi^2$ image, all colors. We also
included a few quantities determined by running the code
\texttt{lephare} on these data: photometric redshift, and ratios of
the minimum $\chi^2$ using galaxy, star, and quasar templates:
$\chi^2_{\rm galaxy}/\chi^2_{\rm star}$, $\chi^2_{\rm
  galaxy}/\chi^2_{\rm quasar}$, and $\chi^2_{\rm quasar}/\chi^2_{\rm
  star}$. Including the transformations of these parameters yields 96
input parameters to the GA-SVM feature selection procedure.

The parameters selected by the GA with occurence larger than
$\mu+\sigma$ times in the posterior distribution are listed in Table
\ref{tab_sg_param}. We also indicate the parameters whose occurence is
larger than $\mu+2\sigma$\,times.  The 15 selected parameters include
\texttt{spread\_model} derived in $g_{\rm P1}$, $r_{\rm P1}$ and
$i_{\rm P1}$, but are dominated by colors (7 of 15). Using the
parameters occuring more than $\mu+2\sigma$ yields similar results,
however with significant overfitting.

\begin{deluxetable}{ccc}
\tabletypesize{\scriptsize}
\tablecaption{Star/Galaxy separation: GA output parameters\label{tab_sg_param}}
\tablewidth{0pt}
\tablehead{
\colhead{Parameter} & \colhead{Transform} & \colhead{Occurence threshold}\tablenotemark{a}}
\startdata
$u_{\rm CFHT}$ & log & \sone\\
$r_{\rm P1}$ & lin & \sone\\
$r_{\rm P1}$ & log & \stwo\\
\texttt{spread\_model\_g} & exp & \sone\\
\texttt{spread\_model\_r} & lin & \stwo\\
\texttt{spread\_model\_r} & log & \stwo\\
\texttt{spread\_model\_i} & exp & \stwo\\
$u_{\rm CFHT}-g_{\rm P1}$ & lin & \sone\\
$u_{\rm CFHT}-z_{\rm P1}$ & lin & \sone\\
$u_{\rm CFHT}-y_{\rm P1}$ & lin& \stwo\\
$g_{\rm P1}-y_{\rm P1}$ & log & \sone\\
$r_{\rm P1}-y_{\rm P1}$ & exp & \sone\\
$i_{\rm P1}-y_{\rm P1}$ & lin & \sone\\
$z_{\rm P1}-y_{\rm P1}$ & exp & \sone\\
$z_{\rm phot}$ & lin & \sone\\
\enddata
\tablenotetext{a}{\sone\,indicates that the parameter occured at least \sone\,times in the posterior distribution of the GA, while \stwo\,indicates that the parameter occured \stwo\,times.}
\end{deluxetable}

In order to quantify the quality of the star/galaxy separation, we use following
definitions for the completeness $c$ and the purity $p$:

\begin{eqnarray}
c_{\rm g} & = & \frac{n_{\rm g}}{n_{\rm g} + m_{\rm g}}\\
p_{\rm g} & =  &\frac{n_{\rm g}}{n_{\rm g} + m_{\rm s}} 
\end{eqnarray}

where $n_{\rm x}$ is the number of objects of class $x$ correctly
classified, and $m_{\rm x}$ is the number of objects of class $x$
misclassified. The same definition holds for the star completeness and
purity.

Based on these definitions, in the case of star/galaxy separation, we
use as cost function:

\begin{equation}\label{eq_chi2_sg}
\begin{split}
  J = \left(\frac{c_{\rm g}-1}{0.005}\right)^2  + \left(\frac{c_{\rm s}-1}{0.005}\right)^2+ \left(\frac{p_{\rm g}-1}{0.005}\right)^2+ \left(\frac{p_{\rm s}-1}{0.005}\right)^2+\\ \left(\frac{f_{\rm SV}-0.1}{0.01}\right)^2.
  \end{split}
\end{equation}

In other words, we choose here to optimize the average completeness and
purity for stars and galaxies, and also penalize high $f_{\rm SV}$,
which amounts to penalize high fractional errors, and overfitting as
well. Other optimization schemes can be adopted.  We perform the
optimization over the only SVM free parameter here, the inverse of
the width of the Gaussian Kernel, $\gamma$. We use a grid search using
a $\log$-spaced binning for $0.01<\gamma<10$.

\begin{figure}[t]
\includegraphics[width=\hsize]{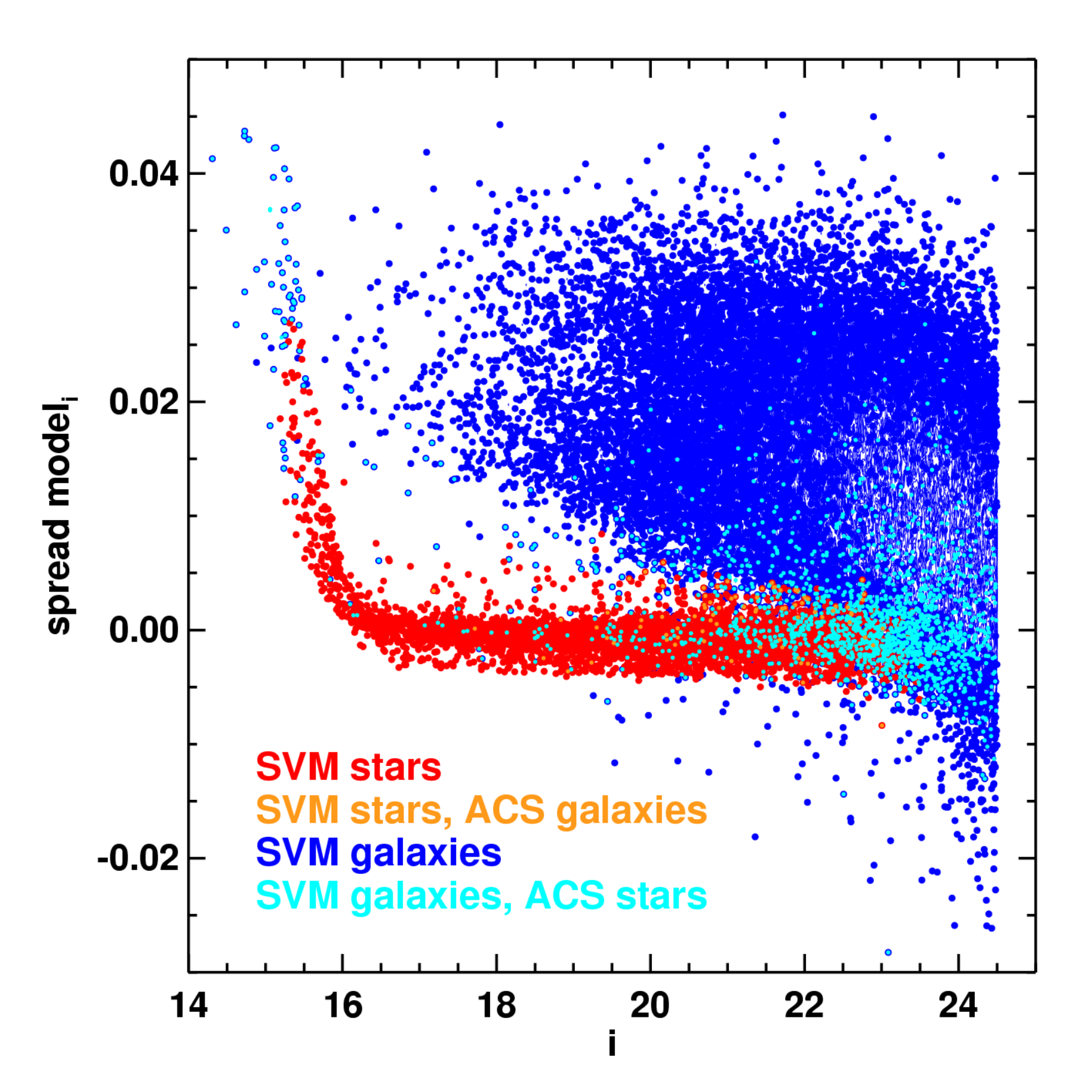}
\figcaption{SExtractor morphological classifier \texttt{spread\_model}
  measured in PS1 $i$ band as a function of $i_{\rm P1}$. We also show
  the result of our baseline classification (training on all objects):
  red points show objects classified as stars by the GA-SVM, and blue
  points objects classified as galaxies. Orange and cyan points show
  misclassified objects: orange are galaxies according to the catalog
  of \citet{Leauthaud_2007} classified as stars, and cyan stars
  classified as galaxies. \label{fig_sg_sep}}
\end{figure}

\begin{figure}[t]
\includegraphics[width=\hsize]{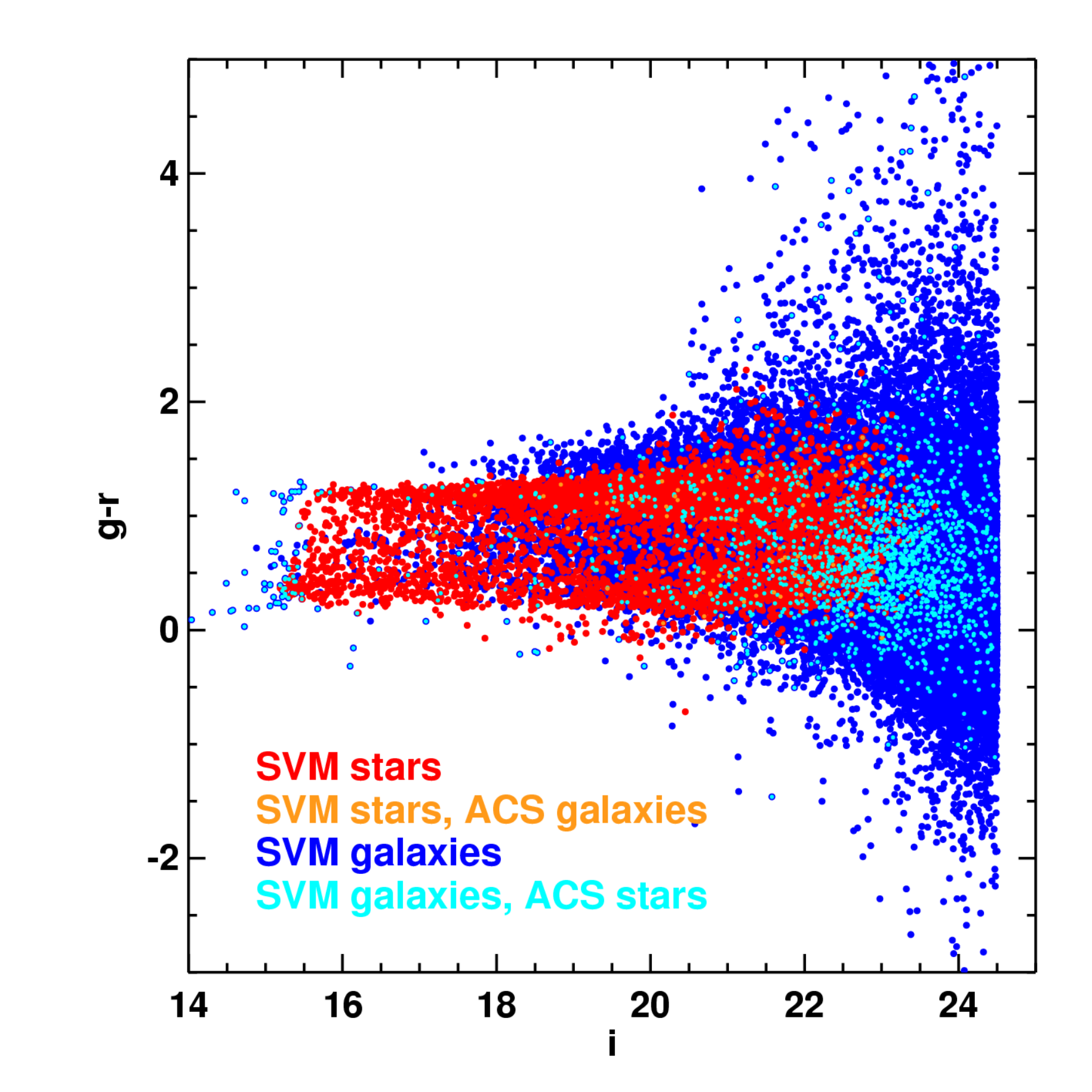} \figcaption{Color magnitude
  diagram $g-i_{\rm P1}$ as a function of $i_{\rm P1}$.  The color coding is the same as in Fig. \ref{fig_sg_sep}\label{fig_sg_sep_color_mag}}
\end{figure}

We show in Fig. \ref{fig_sg_sep} the \texttt{spread\_model} derived in
the PS1 $i$ band, as a function of $i_{\rm P1}$. Fig.
\ref{fig_sg_sep} shows that \texttt{spread\_model} enables to recover
a star sequence down to $i_{\rm P1}\sim 22$. At fainter magnitudes
morphology alone is not able to separate accurately stars from
galaxies at the PS1 angular resolution. The colors coding shows the
result of the GA-SVM classification. We classify objects down to
$i_{\rm P1}=24.5$. We choose this limit because the completeness of
the PS1 data drops significantly beyond 24.5, and also because the
training set we use is valid down to $F814W=25$. Fig. 1 suggests that
the GA-SVM is able to recover the classification at bright magnitudes,
but also to extend it at the faint end. We list in Table
\ref{tab_sg_performance} the percentage of objects classified
correctly.  Our method correctly classifies 97\% of the objects down
to $i_{\rm P1}=24.5$. We can compare our results to those from the PS1
photometric classification server \citep{Saglia_2012}, who used SVM on
bright objects using PS1 photometry. They obtained 84.9\% of stars
correctly classified down to $i_{\rm P1}=21$, and 97\% of galaxies
down to $_{\rm P1}=20$. Our method enables to improve upon those, as
we get 88.6\% of stars correctly classified, and 99.3\% of galaxies
correctly classified in the same magnitude range.

\begin{figure*}
\includegraphics[width=\hsize]{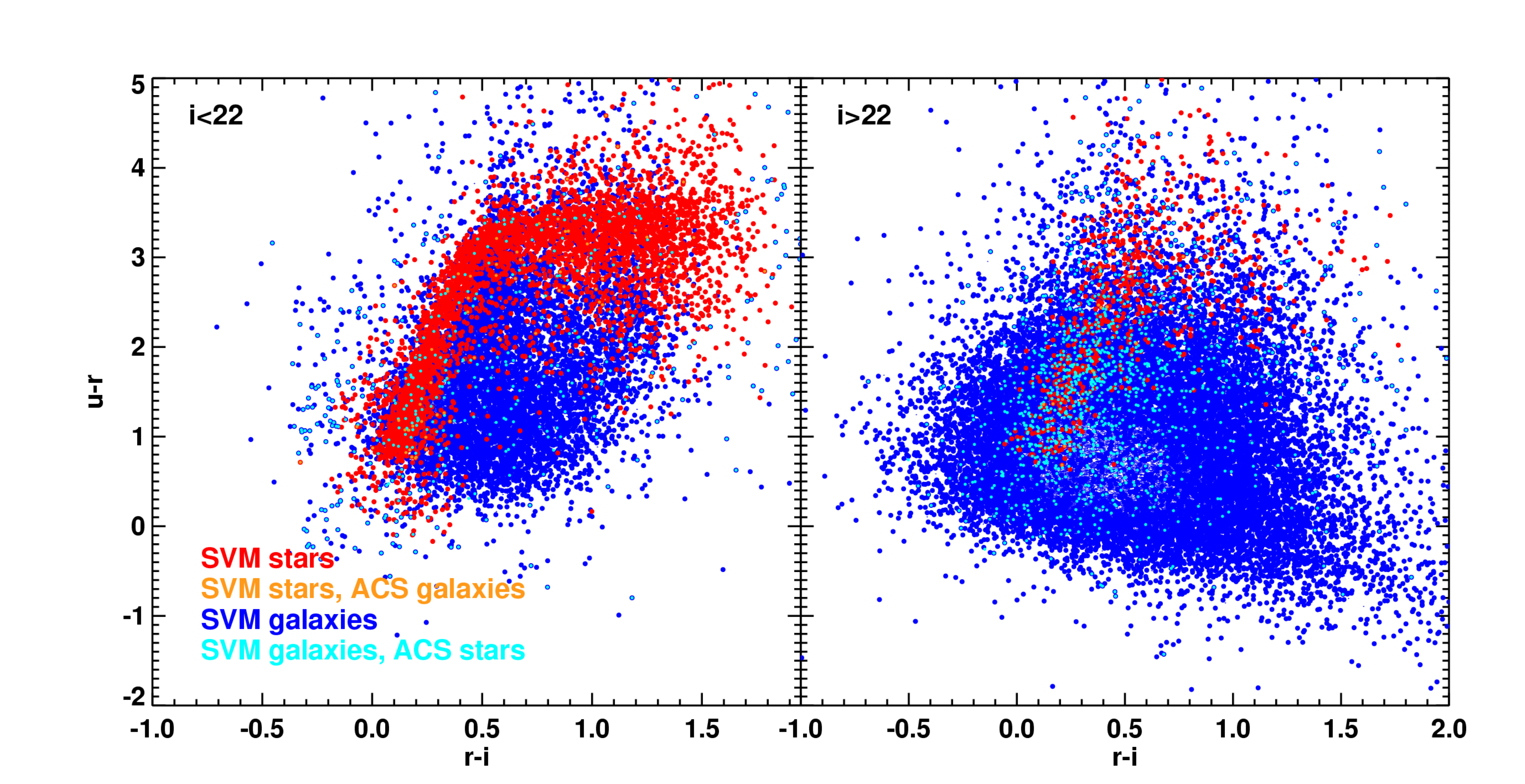} \figcaption{Color-color
  diagram $u-r_{\rm P1}$ as a function of $r-i_{\rm P1}$. The left panel shows objects with $i_{\rm P1}<22$ and the right panel objects with $i_{\rm P1}>22$. The color coding is the same as in Fig. \ref{fig_sg_sep}\label{fig_sg_sep_color_color}}
\end{figure*}

We examine in more details in Fig. \ref{fig_sg_sep_color_mag} and
\ref{fig_sg_sep_color_color} the colors of the objects. These figures
show that the bulk of stars that are misclassified as galaxies are at
the faint end $i\gtrsim 22$, and that these objects are in regions
where the colors of stars and galaxies are similar.  Galaxies
misclassified as stars are brighter $i<22$ but again are in regions
where colors of stars and galaxies overlap. There are also a handful
of very bright stars misclassified as galaxies, in a domain where the
galaxy sampling is very poor. Finally, we classify as galaxies a few
stars showing colors at the outskirt of the color distributions. These
objects might be either misclassified by the ACS photometry, or the
color might be significantly impacted by photometric scatter.

\begin{figure*}[t]
\centering

\includegraphics[scale=0.45]{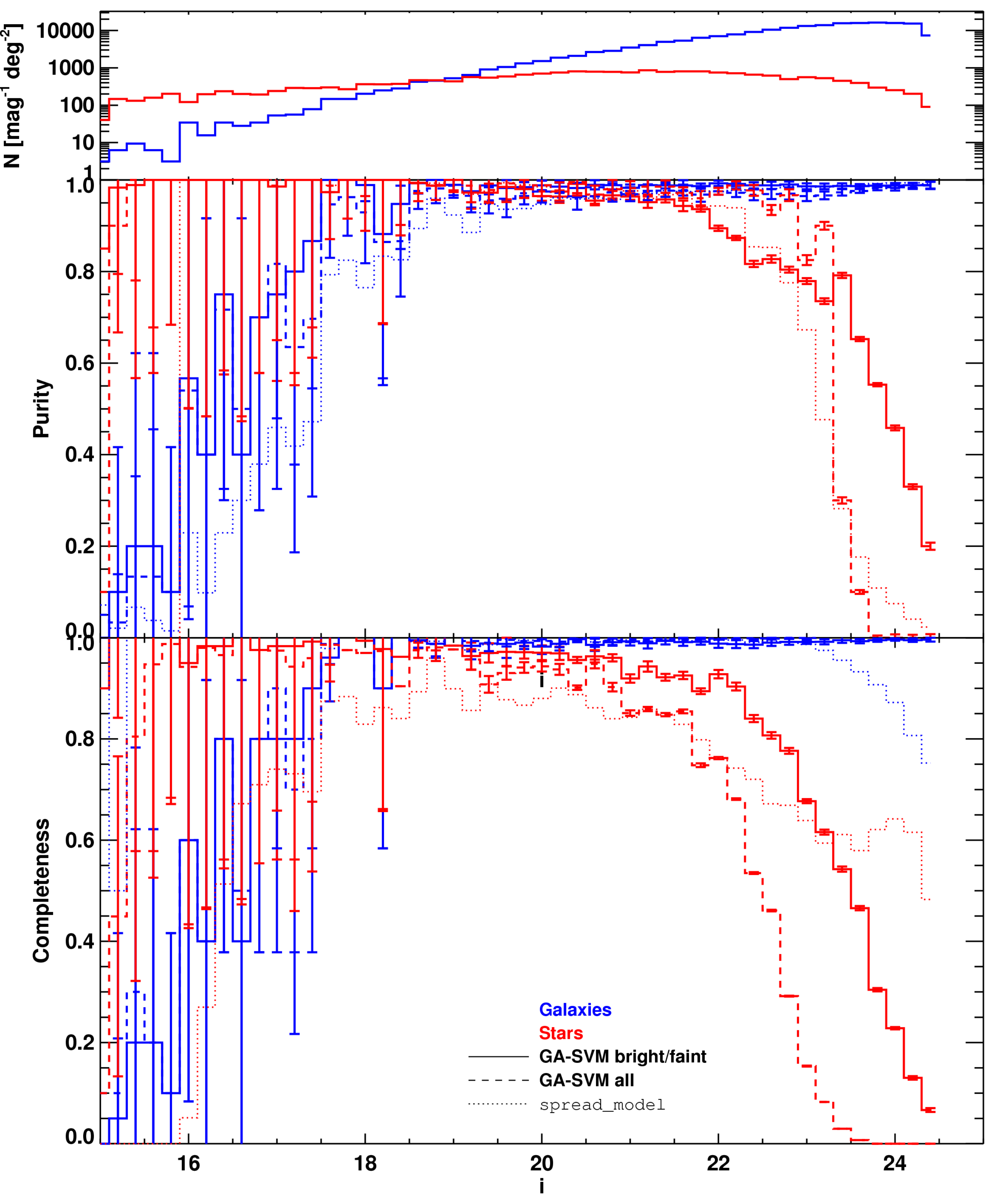}
\figcaption{Quality of the GA-SVM star/galaxy
  classification. \textit{Top}: Galaxy and star
  counts.\textit{Middle:} completeness as a function of PS1 $i$
  magnitude. \textit{Bottom:} Purity as a function of PS1 $i$
  magnitude. On all panels, Red is for stars, and blue for galaxies.
  On the middle and bottom panels, lines show different
  classifications: the dashed lines show the result of our
  classifiction when training with the full sample; solid lines when
  traing on bright ($i_{\rm P1}>22$) and faint ($i_{\rm P1}<22$)
  objects separately; dotted lines show the classification from
  \texttt{spread\_model}.\label{fig_sg_quality}}
\end{figure*}

We show in Fig. \ref{fig_sg_quality} the completeness (left) and
purity (right) of our classifications as a function of $i_{\rm P1}$.

\begin{deluxetable}{cccc}
\tabletypesize{\scriptsize}
\tablecaption{Star/Galaxy separation: performance\label{tab_sg_performance}}
\tablewidth{0pt}
\tablehead{
\colhead{Type}&\colhead{GA-SVM all\tablenotemark{a}} & \colhead{GA-SVM bright/faint\tablenotemark{b}} & \colhead{\texttt{spread\_model}\tablenotemark{c}}}
\startdata
All types & 97.4 & 98.1 & 92.7\\
Galaxies  & 99.8 & 99.2 & 94.5\\
Stars     & 74.5 & 88.5 & 75.9
\enddata
\tablecomments{Percentage of correctly classified objects for each method: \tablenotetext{a}{training and prediction on the full sample} \tablenotetext{b}{Training and prediction on bright and faint objects separately} \tablenotetext{c}{SExtractor \texttt{spread\_model} method}}
\end{deluxetable}

We derive the completeness and purity for each cross-validation
subset, and show in Fig. \ref{fig_sg_quality} the average, and as
error bars the standard deviation.  Most of the features of our
classification seen in Fig. \ref{fig_sg_quality} are due to the fact
that the training sample is unbalanced at the bright end and the faint
end: at the bright end, stars outnumber the galaxies, and the other
way around at the faint end. At the bright end ($i_{\rm P1}<16$, which
is also within the saturation regime), the completeness is higher for
stars than for galaxies, while noisy because of small statistics. Some
bright galaxies are misclassified as stars. At the faint end ($i_{\rm
  P1}>22$), the star completeness decreases, as some stars are
classified as galaxies. The impression given by
Fig. \ref{fig_sg_quality} is striking, as the star completeness
decreases to 0 at $i_{\rm P1}\sim24$. Note however that the stars
represent only 3\% of the overall population at this flux level
\citep{Leauthaud_2007}. The purity shows a similar behaviour at the
bright end. At the faint end however the stars purity is larger than
0.85 for $i_{\rm P1}\lesssim23$. For galaxies, both completeness and
purity are lower than 0.8 at the bright end ($i_{\rm P1}<18$), however
the number of galaxies is small at these magnitudes. At $i_{\rm
  P1}>18$, completeness and purity are independent of magnitude down
to $i_{\rm P1}=24.5$ and larger than 0.95.

We compare this results to the classification obtained with
\texttt{spread\_model} derived in the $i_{\rm P1}$ band. We determine
a single cut in \texttt{spread\_model\_i} using its distribution for
reference stars and galaxies. Our cut is the value of
\texttt{spread\_model\_i} such that
$p(\rm{g|\texttt{spread\_model\_i}}) =
p(\rm{s|\texttt{spread\_model\_i}})$. We show in
Fig. \ref{fig_sg_quality} the completeness and purity obtained with
\texttt{spread\_model\_i} as dotted lines. For $i_{\rm P1}\lesssim
22$, the results from \texttt{spread\_model\_i} and our method are
similar, as at the PS1 resolution, point sources and extended objects
are well discriminated in this magnitude range. At $i_{\rm P1}> 22$,
our baseline method performs better, in particular for galaxies,
which is expected as we add color information. While the
galaxies' purity is similar for both methods, the completeness
obtained with \texttt{spread\_model\_i} drops to 0.75 at $i_{\rm
  P1}=24.5$, but our methods yields a completeness consistent with 1
down to this magnitude. For stars, the purity obtained with the two
methods are similar. On the other hand, the completeness we obtain
drops faster than that obtained with \texttt{spread\_model\_i}. In
order to see whether we can improve our baseline method, we optimize
the SVM parameters independently in two magnitude range: $i_{\rm
  P1}<22$ and $i_{\rm P1}>22$. As galaxies at the faint end outnumber
stars by several orders of magnitude, we add an extra free parameter
for the optimization at $i_{\rm P1}>22$, which attempts at correcting
this sampling issue. In practice we use all stars available, but only
a fraction of the galaxies available, from 1 to 10 times the number of
stars. The results obtained are shown in solid lines on
Fig. \ref{fig_sg_quality}. The results for galaxies are virtually
unchanged compared to our baseline method. For stars, we are able to
improve at the faint end, where the purity is better than that
obtained with \texttt{spread\_model\_i} for $i_{\rm P1}<23$.

As a final check, we also derive the star/galaxy separation by using
SVM only, without selecting the inputs with the GA. The results are
only marginally different. We note however that a parameter space with
lower dimensions is less prone to overfitting with machine learning
methods. On the other hand, even with similar results, a SVM-based
star/galaxy separation with smaller number of inputs parameters is
more likely to generalize properly.

\subsection{Photometric redshifts}\label{sec_photoz}
We used 978 input parameters as inputs for the GA-SVM optimization
procedure: all magnitudes and colors available from the COSMOS dataset
and the transformations of these parameters, as described in
Sect. \ref{sec_optimization}. Hereafter we consider the parameters
that appear at least $\mu +1\sigma$ times in the posterior
distribution, we are left with 131 parameters. The parameters retained
by the GA-SVM are dominated by colors (82\%, 108/131), and in
particular colors involving intermediate and narrow bands (71\%,
93/131). Using the parameters that appear $\mu +2\sigma$ times in the
GA posterior distribution yields 45 parameters, with similar
proportions of colors and intermediate and narrow bands. This is in
line with the conclusions of studies using SED fitting methods which
show that including narrow and intermediate bands improves
significantly the estimation of photometric redshifts
\citep[e.g.][]{Ilbert_2009}.

We quantify the errors on photometric redshifts as $err(z_{\rm spec})
= (z_{\rm phot}-z_{\rm spec})/(1+z_{\rm spec})$, use as a measure of
global accuracy $\sigma$ the normalized median absolute deviation,
defined as $\sigma(z)=1.4826*\textrm{median}(|err(z_{\rm
  spec})-\textrm{median}(err(z_{\rm spec}))|)$.

We define the following cost function for the SVM optimization:
\begin{equation}\label{eq_chi2_zphot}
  J = \left(\frac{\sigma(z)-0.005}{0.0001}\right)^2 + \left(\frac{f_{\rm SV}-0.1}{0.01}\right)^2.
\end{equation}

\begin{figure}[t]
\includegraphics[width=\hsize]{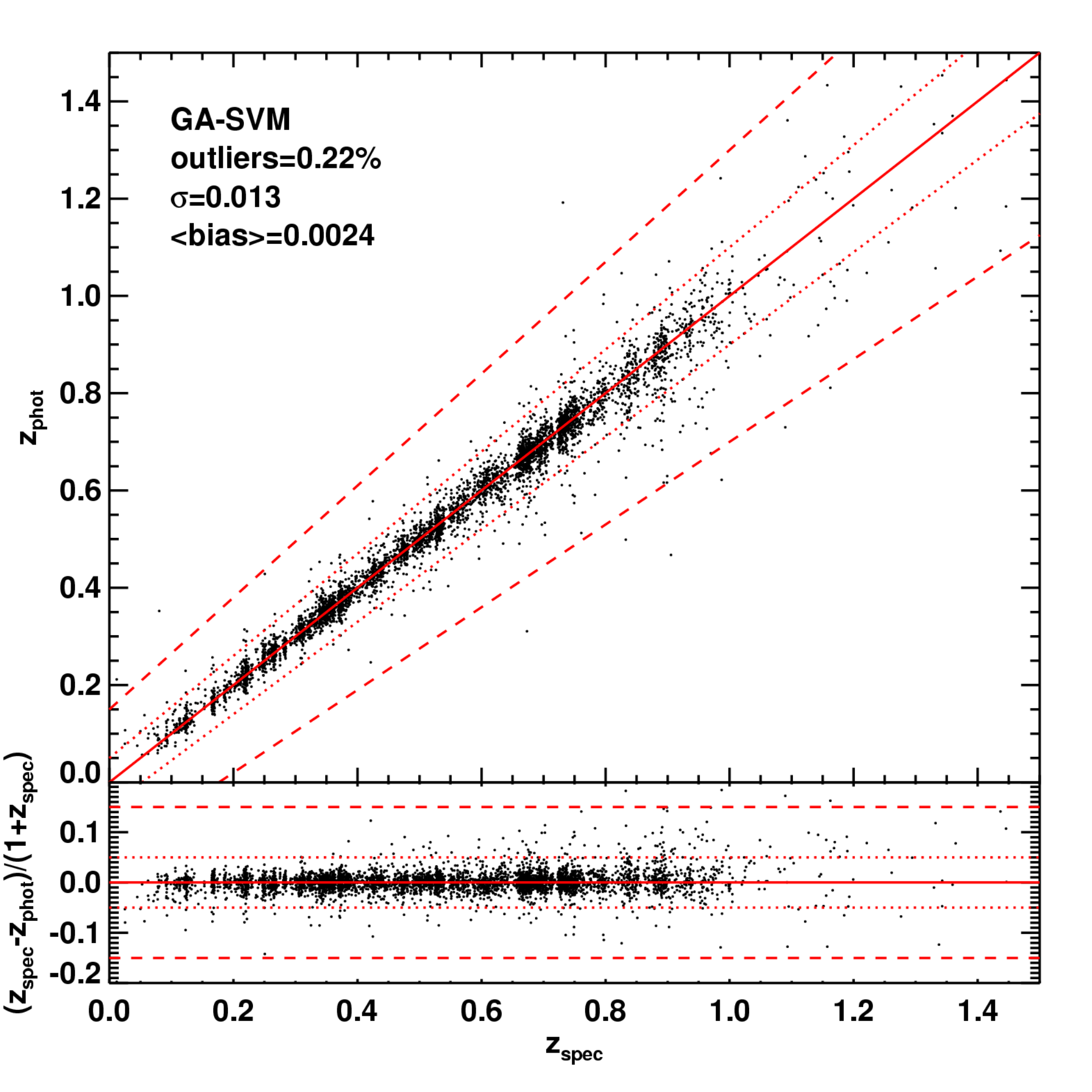}
\figcaption{\textit{Top:} comparison of spectroscopic redshifts with
  photometric redshifts obtained with the GA-SVM. The solid line shows
  identity; the dotted line an error or 0.05 in $1+z$, and the dashed
  line an error of 0.15 in $1+z$. \textit{Bottom:} error of
  photometric redshifts as a function of spectroscopic
  redshifts. Lines are the same as above.\label{fig_zs_zp}}
\end{figure}

We perform the optimization over the 2 SVM free parameters available
here, $\gamma$, and the trade off parameter $C.$ We use a grid search
using a log-spaced binning for $0.01<\gamma<10$ and $0.01<C<500$.

We compare in Fig. \ref{fig_zs_zp} spectroscopic and photometric
redshifts. We obtain an overall accuracy of
0.013\footnote{Using objects with spectroscopic confidence
    class $3\le CC<5$ yields similar results with an accuracy of
    $\sigma(z)=0.015$}. The percentage of outliers, defined as objects with
$|err(z_{\rm spec})| >0.15$, is below 1\%. The average error (bias) is
equivalent to 0; our results do not show any significant bias as a
function of redshift. At high redshifts ($z>1$) the spectroscopic
sampling is small, and so the model is less constrained.  Using the
parameters that appear $\mu +2\sigma$ times in the GA-SVM posterior
distribution yields similar results ($\sigma(z)=0.014$), which shows
that by using 3 times less parameters, the same accuracy can be
achieved.

On a similar sample\footnote{\citet{Ilbert_2009} used an earlier
  version of the catalog we are using here.}, \citet{Ilbert_2009},
using a SED fitting method, obtain an overall accuracy of 0.007. While
our results are slightly worse at face value, we note that we use here
only 2 free parameters for the photometric redshift optimization, and
one model to derive the photometric redshifts (the one from SVM). On
the other hand \citet{Ilbert_2009} rely on 21 SED templates, 30 zero
point offsets (one per band), and an extra parameter describing the
amount $E(B-V)$ of internal dust attenuation. We also explicitly avoid
overfitting, and doing so guarantees the potential for generalization
of these results. Our tests show that we can obtain an overall
accuracy of $\sim 0.01$, however this comes at the price of
significant overfitting (the support vectors are made up from the
whole sample.)

\begin{figure}[t]
\includegraphics[width=\hsize]{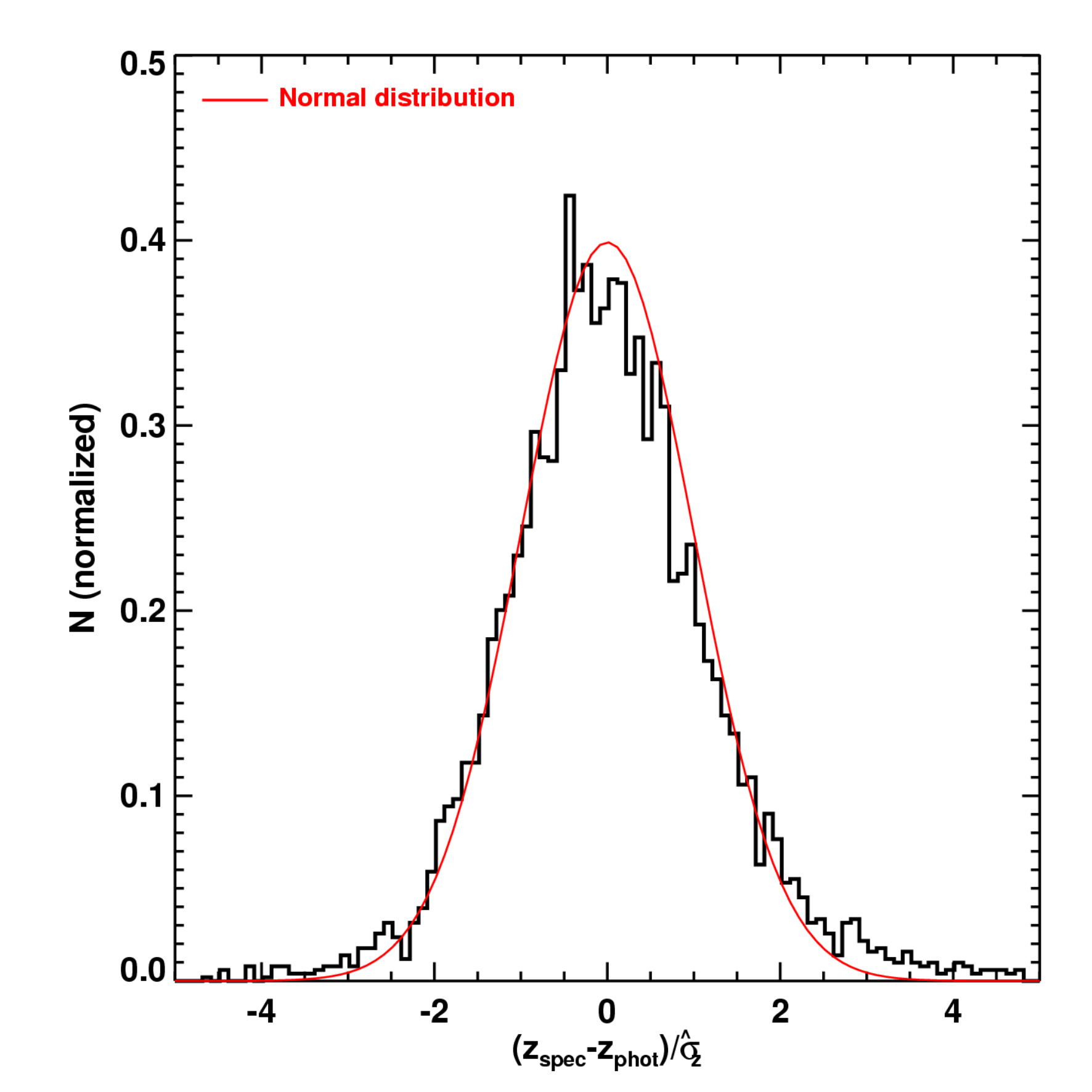}
\figcaption{Empirical estimate of photometric redhift errors. The
  histogram shows the distribution of the ratio $(z_{\rm spec}-z_{\rm
    phot})/\hat{\sigma}_{\rm z}$. The curve is a normal
  distribution.\label{fig_zp_err}}
\end{figure}

As above, we also derive the photometric redshifts by using SVM only,
without selecting the inputs with the GA. In this case the results are
much worse, yielding large errors: $\sigma(z)\sim0.5$. This shows that
the combination of GA and SVM yields better results than SVM alone.

We also test whether we can derive empirical error estimates for each
objects using SVM. We use a variant of the $k-$fold cross validation,
so-called ``inverse''. The usual $k-$fold cross validation consists of
dividing the sample into $k$ subsamples (we use $k=10$ here); for each
subsample predictions are made using the SVM trained on the union of
the other $k-1$ subsamples. To derive error estimates, we use the
inverse $k-$fold cross validation such that we train the SVM on one
subsample, and predict photometric redshifts for the union of the
other $k-1$ subsamples. We have then $k-1$ estimates of $z_{\rm
  phot}$. We derive an empirical error estimate $\hat{\sigma}_{\rm z}$
which is the standard deviation of these estimates. We show in
Fig. \ref{fig_zp_err} the normalized distribution of the ratio
$(z_{\rm spec}-z_{\rm phot})/\hat{\sigma}_{\rm z}$. If our empirical
estimate were an accurate measurement of the actual error, this
distribution should be a Normal one. The comparison with a Normal
distribution shows that this is indeed the case, which suggests that
our error estimates are accurate.

\section{Conclusions}

We present a new combination of two machine learning methods that we
apply to two common problems in astronomy, star/galaxy separation and
photometric redshift estimation. We use Genetic Algorithms to select
relevant features, and Support Vector Machines to estimate the
quantity of interest using the selected features. We show that the
combination of these two methods yields remarkable results, and offers
an interesting opportunity for future large surveys which will gather
large amount of data. In the case of star/galaxy separation,
  the improvements over existing methods are a consequence of adding
  more information, while for photometric redshifts, it is rather the
  selection of the input information fed to the machine learning
  methods. This shows that the combination of GA and SVM is very
  efficient in the case of problems with large dimensions.

We first apply the GA-SVM method to star/galaxy separation in the PS1
Medium Deep Survey. Our baseline method correctly classifies 97\% of
objects, in particular virtually all galaxies. Our results improve
upon the new SExtractor morphological classifier,
\texttt{spread\_model}, which is expected as we added color
  information compared to morphology only. We show how these results
can be further improved for stars by training separately bright and
faint objects, and taking into account the respective number of stars
and galaxies to avoid being dominated by one population.

We then apply the GA-SVM method to photometric redshift estimation
for the zCOSMOS bright sample. We obtain an accuracy of 0.013, which
compares well with results from SED fitting, as we are using only 2
free parameters. We also show that we can derive accurate error
estimates for the photometric redshifts.

We present here a proof-of-concept of a new method that can be
modified or improved depending on the problem at stake. For instance,
one can substitute another machine learning tool to SVM (such as
Random Forests for instance) to derive the quantity of
interest. Furthermore the criterion used to select the final number of
features from the GA posterior distribution can be also be optimized
beyond the one we use here. All these tools will enable to use as much
information as possible in an efficient way for future large surveys.

\acknowledgements

The Pan-STARRS1 Surveys (PS1) have been made possible through
contributions of the Institute for Astronomy, the University of
Hawaii, the Pan-STARRS Project Office, the Max-Planck Society and its
participating institutes, the Max Planck Institute for Astronomy,
Heidelberg and the Max Planck Institute for Extraterrestrial Physics,
Garching, The Johns Hopkins University, Durham University, the
University of Edinburgh, Queen's University Belfast, the
Harvard-Smithsonian Center for Astrophysics, the Las Cumbres
Observatory Global Telescope Network Incorporated, the National
Central University of Taiwan, the Space Telescope Science Institute,
the National Aeronautics and Space Administration under Grant
No. NNX08AR22G issued through the Planetary Science Division of the
NASA Science Mission Directorate, the National Science Foundation
under Grant No. AST-1238877, the University of Maryland, and Eotvos
Lorand University (ELTE) and the Los Alamos National Laboratory.

\appendix

\section{Support Vector Machines: short description}\label{app_svm}
We provide here a short description of the SVM. More detailed
presentations of the formalism can be found elsewhere
\citep[e.g.][]{Smola_1998, Vapnik_1995, Vapnik_1998}. For the sake of
brevity, we present here only the equations relevant to regression
with SVM. Equations for classification are similar, except a few
differences that we mention whenever necessary.

The training data usually consists of a number objects with input
parameters, $\vec{x}$, of any dimension, and the known values of the
quantity of interest, $\vec{y}$.  The goal of SVM is to find a
function $f$ such as
\begin{equation}
  f = \vec{w}.\vec{x} + b
\end{equation}
which yields $\vec{y}$ with a maximal error $\epsilon$, and such as
$f$ is as flat as possible. In other words, the amplitude of the slope
$\vec{w}$ has to be minimal. One way to achieve this is to minimize
the norm $\frac{1}{2}|w|^2$ with the condition $|y_i -
\vec{w}.\vec{x_i} -b|\le \epsilon$. The margin in that case is
$\frac{2}{|w|}$. In other words, SVM attempt to regress with the
largest possible margin. 

In a number of problems, the data can not be separated using a fix,
hard margin. It is then useful to allow some points to be
misclassified.  One uses a "'soft margin'', which enables
to allow some errors in the results. The modified minimization reads:

\begin{align}
  \textrm{minimize} & \left.\begin{array}{l}\frac{1}{2} |w|^2 + C\sum_i \xi_i +\xi_i^* \label{eq_svm_min}\\ \end{array}\right. \\
  \textrm{subject to}: & \left\{\begin{array}{l} y_i - w.x_i - b \le \epsilon + \xi_i \\
                    -y_i + w.x_i + b \le -\epsilon + \xi_i^* \\
                     \xi_i \textrm{,} \xi_i^* \ge 0
  \end{array}\right.
  \end{align}

where $\xi_i, \xi_i^*$ are "slack variables'', and $C$ is a free
parameter which controls the soft margin. The larger $C$, the harder
the margin: larger error are penalized.

Using Lagrange multiplier analysis, it can be shown that the slope $w$ can be written as:
\begin{equation}\label{eq_sv}
  w = \sum_{i=1}^{n} (\alpha_i - \alpha_i^*)x_i
\end{equation}
where $\alpha_i, \alpha_i^*$ are the Lagrange multipliers, which
satisfy $\sum_{i=1}^{n} \alpha_i - \alpha_i^* = 0$, and $\alpha_i, \alpha_i^* \in [0,C]$.

Eq. \ref{eq_sv} shows that the solution of the minimization problem is
a linear combination of a number of input data points. In other words,
the solution is based on a number of support vectors, the number of
training samples where $\alpha_i - \alpha_i^* \ne 0$.

The above equations use the actual values of the data, assuming that
the separation can be performed linearly. For most high dimension
problems, this assumption is not valid any more. The fact that only a
scalar product between the support vectors and the input data is
required enables to use the so-called "kernel trick''. The idea behind
the trick is that one can use functions which satisfy a number of
conditions to map the input space to another where the separation can
be performed linearly.  Eq. \ref{eq_sv} then becomes:

\begin{equation}\label{eq_sv_kernel}
  w = \sum_{i=1}^{n} (\alpha_i - \alpha_i^*)\Phi(x_i)
\end{equation}

where the kernel $k(x,x')=\langle\Phi(x)\Phi(x')\rangle$.

Finally, a slightly modified version of the algorithm ($\nu$SVR)
allows to determine $\epsilon$ and control the number of support
vector.  A parameter $\nu\in [0,1]$ is introduced such that
Eq. \ref{eq_svm_min} becomes

\begin{equation}
  \textrm{minimize}\quad \frac{1}{2} |w|^2 + C(\nu\epsilon +\sum_i \xi_i +\xi_i^*). \label{eq_nusvm_min}
\end{equation}

It can be shown that $\nu$ is the upper limit on the fraction of
errors, and the lower limit of the fraction of support vectors.

In the regression case described here, the free parameters for
$\nu$SVR are the trade off parameter $C$ and all the kernel parameters
(assuming that one fixes $\nu$, which allows control of the error and
the fraction of support vectors). In the classification case
($\nu$SVC), the free parameters are only those from the kernel which
is used, as $\nu$ replaces the trade off parameter $C$, and again, one
would usually fix $\nu$.

\end{document}